\documentclass[12pt,preprint]{aastex}
\usepackage{rotate}

\slugcomment{Accepted for publication on  ApJ Letters}

\shorttitle{ACS bursts from 1E\,1547.0--5408}

 \shortauthors{S.~Mereghetti~et~ al.}

\def\approxgt{\mathrel{\hbox{\rlap{\lower.55ex \hbox {$\sim$}}
        \kern-.3em \raise.4ex \hbox{$>$}}}}
\def\approxlt{\mathrel{\hbox{\rlap{\lower.55ex \hbox {$\sim$}}
        \kern-.3em \raise.4ex \hbox{$<$}}}}

\def\pdot {\dot P}

\def\ltsima{$\; \buildrel < \over \sim \;$}
\def\lsim{\lower.5ex\hbox{\ltsima}}
\def\gtsima{$\; \buildrel > \over \sim \;$}
\def\gsim{\lower.5ex\hbox{\gtsima}}

\def\sgr {1E\,1547.0--5408}

\begin{document}


\title{Strong bursts from the anomalous X-ray pulsar \sgr\ observed with the \emph{INTEGRAL}/SPI Anti-Coincidence Shield}

\author{S. Mereghetti\altaffilmark{1}, D. G\"{o}tz\altaffilmark{2}, G. Weidenspointner\altaffilmark{3,9}, A. von Kienlin\altaffilmark{3}, P. Esposito\altaffilmark{1,4}, A. Tiengo\altaffilmark{1}, G. Vianello\altaffilmark{1}, G.~L. Israel\altaffilmark{5}, L. Stella\altaffilmark{5}, R. Turolla\altaffilmark{6,7}, N. Rea\altaffilmark{8}, and S. Zane\altaffilmark{7}}
\affil{$^{1}$ INAF - Istituto di Astrofisica Spaziale e Fisica Cosmica Milano, via E.~Bassini 15, I-20133 Milano, Italy; \url{sandro@iasf-milano.inaf.it}}
\affil{$^{2}$ CEA Saclay, DSM/IRFU/Service d'Astrophysique, Orme des Merisiers, B\^at. 709, F-91191 Gif-sur-Yvette, France}
\affil{$^{3}$ Max-Planck-Institut f\"{u}r extraterrestrische Physik,
Giessenbachstrasse,  Postfach 1312, D-85741 Garching, Germany}
\affil{$^{4}$ INFN - Istituto Nazionale di Fisica Nucleare, Sezione di Pavia, via A.~Bassi 6, 27100 Pavia, Italy}
\affil{$^{5}$ INAF - Osservatorio Astronomico di Roma,
 via Frascati 33, I-00040 Monteporzio Catone, Italy}
\affil{$^{6}$ Universit\`a di Padova, Dipartimento di Fisica, via
Marzolo 8, I-35131 Padova, Italy}
 \affil{$^{7}$ Mullard Space
Science Laboratory, University College London, Holmbury St. Mary,
Dorking Surrey, RH5 6NT, United Kingdom}
 \affil{$^{8}$
Astronomical Institute ``Anton Pannekoek'', University of
Amsterdam, Kruislaan 403, 1098SJ, Amsterdam, The Netherlands}
 \affil{$^{9}$ MPI Halbleiterlabor, Otto-Hahn-Ring 6, 81739 Muenchen, Germany }

\begin{abstract}
In January 2009, multiple short bursts of soft gamma-rays were
detected from the direction of the anomalous X-ray pulsar \sgr\ by
different satellites. Here we report on the observations obtained
with the \emph{INTEGRAL} SPI-ACS detector during the period with
the strongest bursting activity. More than 200 bursts were
detected at energies  above 80 keV in a few hours on January 22.
Among these, two remarkably bright events showed pulsating tails
lasting several seconds and modulated at the 2.1 s spin period of
\sgr. The energy released in the brightest of these bursts was of
a few 10$^{43}$ erg, for an assumed distance of 10 kpc. This is
smaller than that of the three  giant flares seen from soft
gamma-ray repeaters, but higher than that of typical bursts from
soft gamma-ray repeaters and anomalous X-ray pulsars.
\end{abstract}


\section{Introduction}

The soft gamma-ray repeaters (SGRs) are  a small group of
high-energy sources that emit short ($<$1 s) bursts of soft
gamma-rays with peak luminosities up to 10$^{41}$ erg s$^{-1}$
during sporadic periods of activity. Much more rarely they emit
giant flares, releasing up to 10$^{46}$ ergs. SGRs are thought to
be magnetars, i.e. isolated neutron stars whose persistent and
bursting emission is powered by extremely high magnetic fields,
$B>10^{14}$--10$^{15}$ G \citep{tho95,tho96}. In recent years
bursts similar to those of SGRs have been detected also from
members of another class of sources, the anomalous X-ray pulsars
(AXPs) \citep{gav02,kas03,isr08}. Evidence is accumulating that
there might not be any physical reason to distinguish among these
two groups of neutron stars, with their different classification
reflecting only the way they were originally discovered. A recent
review of AXPs and SGRs is given in \citet{mer08}.

The source discussed here, \sgr , was discovered almost 30 years
ago \citep{lam81}, but it attracted little attention until it was
proposed as a possible AXP  on the basis of new X-ray and optical
studies \citep{gel07}. The source is transient, and it is located
in the supernova remnant G\,327.24--0.13.  The discovery of radio
pulsations with period $P=2.1$ s  and  period derivative
$\pdot=2.3\times10^{-11}$ s s$^{-1}$ confirmed the AXP
classification \citep{cam07c}.

In early 2008 October \sgr\ showed  enhanced X-ray activity, with
the emission of several short bursts  in a few days, accompanied
by an increase in its persistent X-ray flux (Israel et al., in
preparation). No further bursts were reported after that until the
source started a new period of  strong activity on 2009 January
22, as testified by the numerous bursts detected with \emph{Swift}
\citep{GCN8833}, \emph{Fermi}/GBM \citep{GCN8835,GCN8838},
\emph{INTEGRAL} \citep{GCN8837,GCN8841}, \emph{Suzaku}
\citep{GCN8845}, Konus-\emph{Wind} \citep{GCN8851} and
\emph{RHESSI} \citep{GCN8857}. Here we report on observations
obtained with the SPI-ACS instrument on board the \emph{INTEGRAL}
satellite that provided   uninterrupted monitoring of the most
active bursting period, thanks to its highly elliptical orbit.

\begin{figure}
\includegraphics[angle=00,width=15cm]{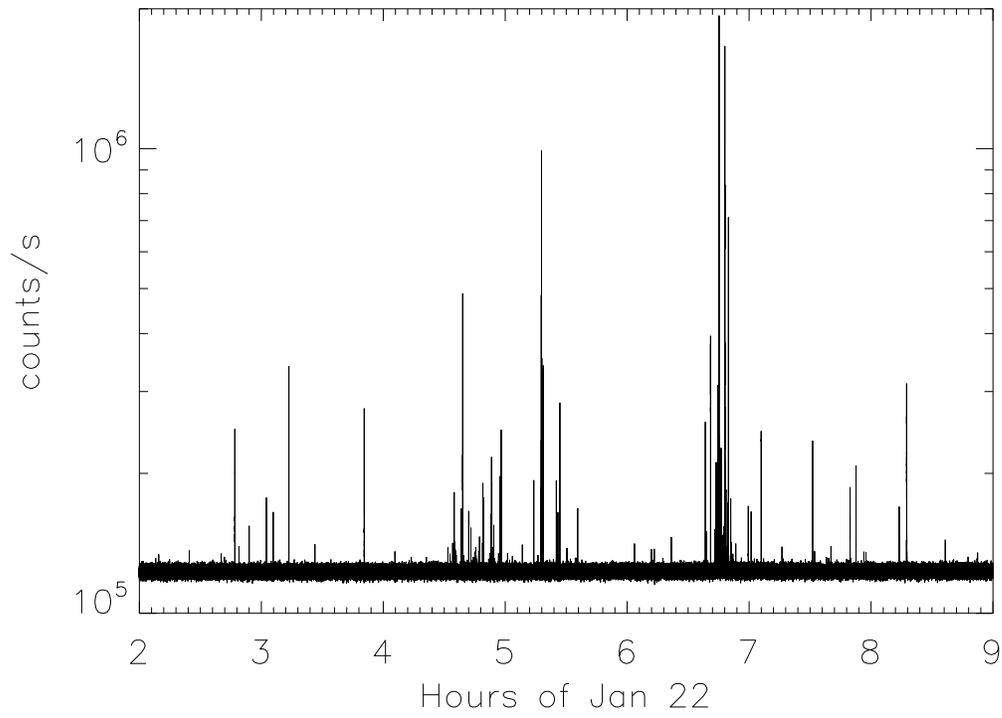}
\caption{\label{lc29} Light curve of \sgr\ obtained with the
SPI-ACS during the most active period on 2009 January 22. The time
binning is 50 ms. }
\end{figure}

\begin{figure}
\includegraphics[angle=90,width=15cm]{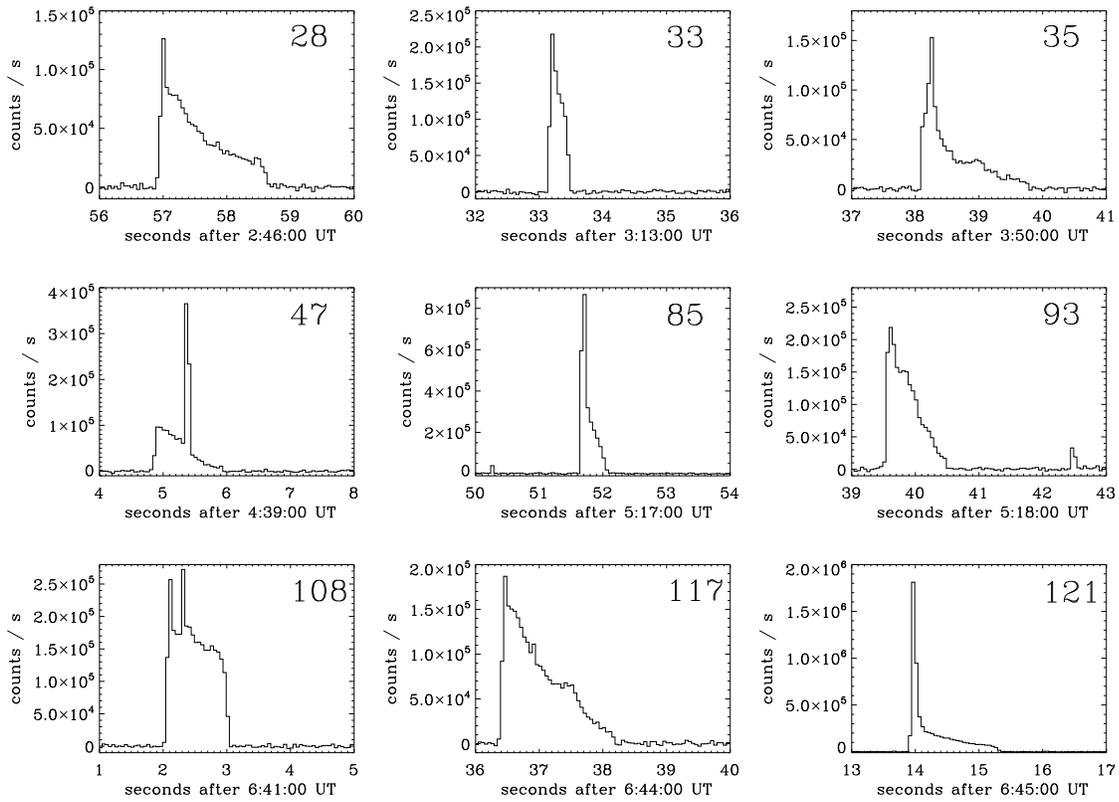}
\caption{\label{figurine} SPI-ACS light curves of a few bursts
observed on 2009 January 22. See Table~\ref{bursts} for their
properties.}
\end{figure}

\begin{figure}
\includegraphics[angle=0,width=16cm]{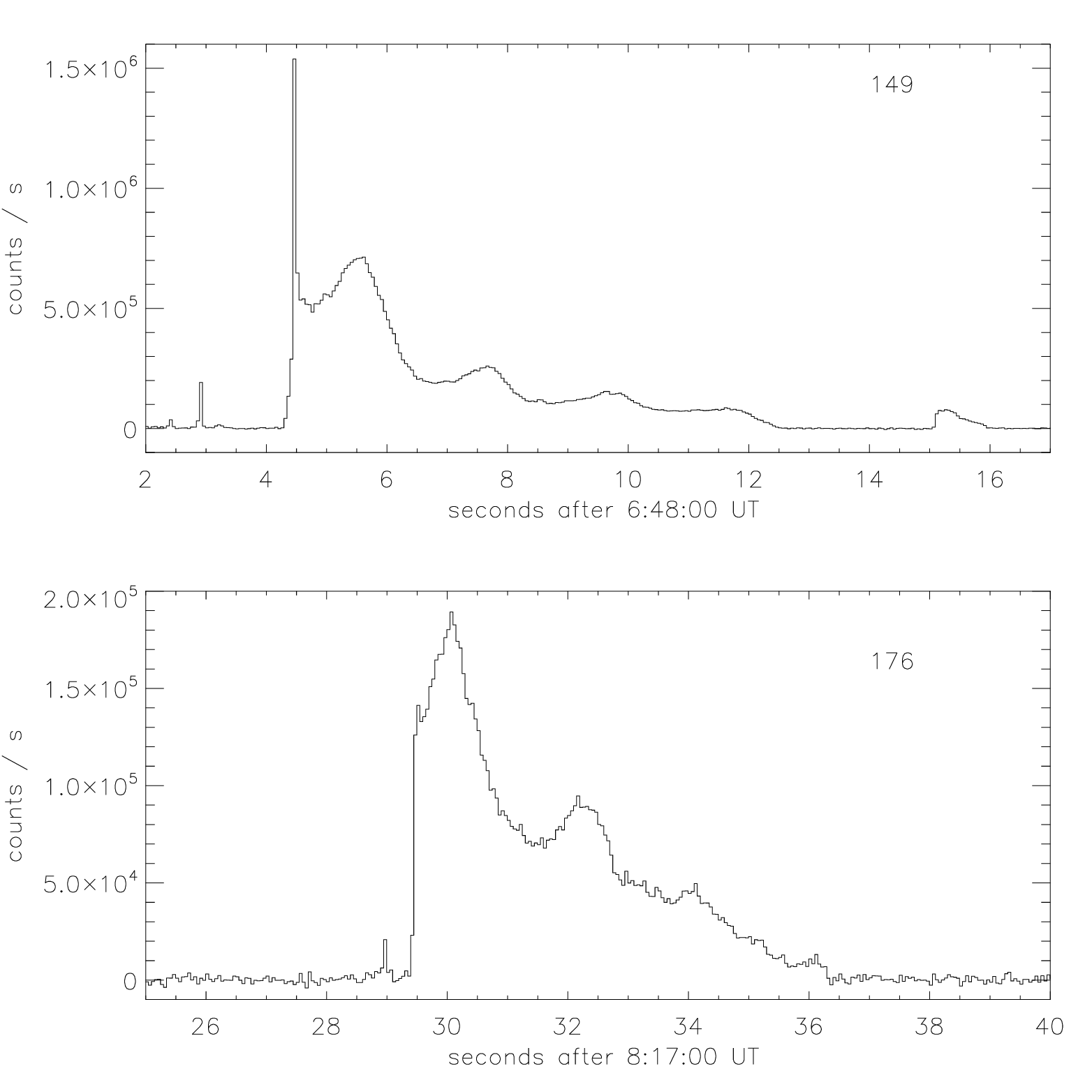}
\caption{\label{lc2} SPI-ACS light curves of the two longest
bursts seen from \sgr\ on 2009 January 22. Note that the initial
spike of the 6:48:04 UT burst was probably brighter than shown in
the figure, due to the non negligible instrumental dead-time at
high count rates. }
\end{figure}

\section{Results}

The Anti-Coincidence Shield (ACS) of the SPI instrument
\citep{ved03}, besides serving to veto the background in the
germanium spectrometer, is routinely used as a nearly
omni-directional detector for gamma-ray bursts \citep{von03}. It
consists of  91 bismuth germanate (BGO) scintillator crystals of
thickness between 16 and 50 mm that provide a large effective area
at $E>80$ keV for most directions. The data used here consist of
the overall light curve resulting from the OR-ed veto signals of
all the ACS crystals. The time bin is 50 ms, and no energy and
directional information is available.

During the burst observations of 2009 January 22,  the
\emph{INTEGRAL} pointing direction varied by less than $\sim$
10\degr. The change in the ACS effective area resulting from the
varying instrumental direction is negligible. We therefore used
for the whole analysis the instrumental response computed for the
zenith and azimuth of \sgr\ at the time of the   event with the
largest fluence (burst n.~149, see below). These are
$\theta=61\degr$ (from the SPI pointing axis)  and $\phi=322\degr$
($\phi=0\degr$ corresponds to the satellite Sun-pointing side,
i.e. the direction from SPI toward the IBIS instrument). The ACS
effective area as a function of energy was computed with Monte
Carlo simulations based on a detailed mass modelling of the SPI
spectrometer and surrounding material, including the satellite
structure and other instruments \citep{stu03}. These simulations
were carried out using MGGPOD \citep{wei05}, a suite of Monte
Carlo tools developed for modelling high energy astronomy
instruments. In a similar way we computed the ACS effective area
for the satellite orientation corresponding to the bright burst of
January 25, which was observed close to the on-axis direction
($\theta=5\degr$).

The light curve covering the most active period is shown in
Fig.~\ref{lc29}. During the observation of the bursts, the
background count rate in the ACS was stable,  within 1\%,  at
$1.23\times10^5$ counts s$^{-1}$. Adopting a flux threshold at
5$\sigma$ above the background, 233 bursts were detected  from
18:11 UT of January 21 to 4:27 UT of January 23. Their peak
fluxes, over a time integration interval of 50 ms, cover the range
from $\sim$ 10$^{4}$ to more than 10$^{6}$ counts
s$^{-1}$,\footnote{While dead time and saturation effects in the
ACS crystals and electronics are negligible ($<$1\%) below a few
10$^5$ counts s$^{-1}$, the results   for fluxes above this level
should be considered as lower limits.} while the fluences span the
range from $\sim$ 500 to $\sim$ $2\times10^{6}$ counts. We derived
the distributions of peak fluxes and fluences and fitted them
using the Maximum Likelihood method \citep{crawford70}. Assuming a
power-law distribution with index $\alpha$ for the integral
distributions ($N(>$$S)\propto S^{-\alpha}$), we obtained $\alpha
= 1.03\pm0.08$ for the peak flux distribution, and  $\alpha =
0.75\pm0.06$ for the fluence distribution.

The time distribution of the bursts is strongly non-uniform.  Most
of them  are clustered in two time intervals, from 4:34 to 5:28 UT
and from 6:43 to 6:51 UT on January 22. The total fluence of all
the bursts detected in these two time intervals are of $\sim$
$1.5\times10^{6}$ and $\sim$ $2.8\times10^{6}$  ACS counts,
respectively. We computed the phases of the onset time of each
burst with respect to the neutron-star rotation, based on a
phase-connected timing solution we derived from \emph{Swift}/XRT
data. We found that the burst onset times are not correlated with
the neutron-star rotation phase.

The light curves of a few bursts, including that of the brightest
one (burst n.~121, at 06:45:14 UT), are shown in
Fig.~\ref{figurine}. The properties of the brightest events  are
given in Table~\ref{bursts}. Two bursts had durations longer than
the spin period of \sgr\ and clearly showed a modulation at 2.1 s
(see Fig.~\ref{lc2}). They occurred at 6:48:04 UT (burst n.~149)
and 8:17:29 UT (burst n.~176). The first burst was brighter, it
started with a very bright and short initial spike ($\sim$0.3 s)
followed by a $\sim$8 s long pulsating tail. About 2.5 s later a
second $\sim$0.9 s long burst (burst n.150) was observed, whose
profile is not in phase with the preceding pulses.

To estimate the energetics of these events we must assume a
spectral shape, since no energy resolution is provided by the ACS
detectors. SGR bursts are well fit in the hard X-ray range by
thermal bremsstrahlung spectra with temperatures of $\sim$20-40
keV. The bright burst of January 25 (see last line of
Table~\ref{bursts}) was in the field of view of the IBIS
instrument \citep{ube03}. We could therefore analyze its  15-300
keV spectrum measured with the ISGRI detector \citep{leb03},
obtaining a bremsstrahlung temperature of kT=46 keV \footnote{Due
to the burst brightness, which saturated the ISGRI telemetry, it
was not possible to obtain also a measure of the peak flux and
fluence.}. We therefore adopt in the following an optically thin
thermal bremsstrahlung with $kT=40$ keV to convert the ACS count
rates. With this assumption, the fluence (25 keV--2 MeV) of the
pulsed tail of burst n.~149 is $\sim$ $2.5\times10^{-4}$ erg
cm$^{-2}$. Extrapolating this spectrum to lower energy, and
assuming isotropic emission, we derive a total energy (at $E>1$
keV) in the pulsating tail of $\sim$ $2.4\times10^{43}\
d^{2}_{10\,\mathrm{kpc}}$ erg. Note that this is a conservative
estimate, since a lower temperature would result in a higher
luminosity.  The corresponding values for the second pulsed burst
(n.~176), which was slightly shorter,  are a factor $\sim$ 4
smaller.

The initial spike of  burst n.~149 reached a peak count rate of
$1.5\times10^6$ counts s$^{-1}$ ($\Delta T=50$ ms) and was
surpassed in intensity only by  burst n.~121 that occurred about
three minutes earlier. In both cases, only lower limits to the
flux can be derived, due to the non-linearity of the ACS response
at such high count rates. The latter are reported in
Table~\ref{bursts}, again for an assumed $kT=40$ keV thermal
bremsstrahlung spectrum. However, it is likely that these bursts
had significantly harder spectra. In fact, the initial spikes
observed in giant flares from other three SGRs had spectra
significantly harder than those of the pulsating tails. Evidence
for harder spectra was also reported in other bright bursts from
\sgr\ observed with Konus/\emph{Wind} on January 25 and 29
\citep{GCN8858,GCN8863}. For example, an exponentially cut-off
power law with photon index $\Gamma=-1$ and $E_{\mathrm{cut}}=400$
keV,   results in 25 keV--2 MeV fluxes and fluences about 30\%
smaller than those reported in Table~\ref{bursts}. Extrapolating
this spectrum,  gives lower limits to the peak luminosity of 6 and
$5 \times10^{42}\ d^{2}_{10\,\mathrm{kpc}}$ erg s$^{-1}$ ($E> 1$
keV) for bursts n.~121 and n.~149, respectively.

\begin{table*}[htbp!]
\begin{center}
\caption{\label{bursts}Bursts with peak flux higher than $10^{5}$
ACS counts s$^{-1}$. }
\begin{tabular}{llcrrl}
\hline \hline
ID &  Start UT & Duration   & Peak flux$^{(a)}$  & Fluence$^{(a)}$   & Notes \\
   &  Jan 22   & (s)    &  10$^{-5}$ erg cm$^{-2}$ s$^{-1}$ & 10$^{-5}$ erg cm$^{-2}$ & \\
\hline
 28  &  2 46 57.0  &  1.70  &  $1.83\pm0.04$  &  $1.101\pm0.011$  & \\
 33  &  3 13 33.2  &  0.35  &  $3.16\pm0.05$  &  $0.643\pm0.006$  & \\
 35  &  3 50 38.1  &  1.30  &  $2.22\pm0.04$  &  $0.787\pm0.010$  & \\
 47  &  4 39 04.9  &  0.95  &  $5.30\pm0.05$  &  $1.099\pm0.009$  & \\
 71  &  4 57 29.8  &  0.65  &  $1.08\pm0.04$  &  $0.381\pm0.007$  & \\
 73  &  4 58 03.0  &  0.20  &  $1.82\pm0.04$  &  $0.234\pm0.004$  & \\
 80  &  5 17 44.1  &  0.15  &  $5.23\pm0.05$  &  $0.483\pm0.004$  & \\
 82  &  5 17 48.2  &  0.55  &  $1.63\pm0.04$  &  $0.521\pm0.007$  & \\
 85  &  5 17 51.7  &  0.45  & $12.57\pm0.07$  &  $1.914\pm0.008$  & \\
 89  &  5 18 01.7  &  0.20  &  $3.35\pm0.05$  &  $0.339\pm0.004$  & \\
 92  &  5 18 32.9  &  0.10  &  $2.97\pm0.05$  &  $0.221\pm0.003$  & \\
 93  &  5 18 39.5  &  1.00  &  $3.17\pm0.05$  &  $1.442\pm0.009$  & \\
 98  &  5 26 52.9  &  0.25  &  $2.34\pm0.04$  &  $0.359\pm0.005$  & \\
106  &  6 38 27.9  &  0.25  &  $1.96\pm0.04$  &  $0.265\pm0.004$  & \\
108  &  6 41 02.1  &  1.00  &  $3.95\pm0.05$  &  $2.354\pm0.010$  & \\
117  &  6 44 36.4  &  1.75  &  $2.71\pm0.05$  &  $1.928\pm0.012$  & \\
119  &  6 45 00.2  &  0.20  &  $2.20\pm0.04$  &  $0.213\pm0.004$  & \\
120  &  6 45 12.3  &  0.20  &  $1.72\pm0.04$  &  $0.184\pm0.004$  & \\
121  &  6 45 13.9  &  1.45  & $>$$26.27\pm0.09$ &  $>4.587\pm0.013$  & brightest \\
130  &  6 46 14.8  &  0.15  &  $1.51\pm0.04$  &  $0.106\pm0.003$  & \\
141  &  6 47 57.1  &  0.35  & $13.10\pm0.07$  &  $1.819\pm0.007$  & \\
147  &  6 48 02.9  &  0.15  &  $2.78\pm0.05$  &  $0.169\pm0.003$  & \\
149  &  6 48 04.3  &  8.15  &$>$$22.31\pm0.09$& $>27.759\pm0.031$  & total \\
149-s&  6 48 04.3  &  0.30  &$>$$22.31\pm0.09$ &  $>2.308\pm0.008$ & initial spike\\
149-t&  6 48 04.6  &  7.85  & $10.34\pm0.06$  & $25.451\pm0.030$  & pulsed tail \\
150  &  6 48 15.1  &  0.85  &  $1.13\pm0.04$  &  $0.584\pm0.008$  & \\
156  &  6 49 49.0  &  0.25  &  $8.54\pm0.06$  &  $0.826\pm0.005$  & \\
164  &  7  5 56.8  &  0.35  &  $1.80\pm0.04$  &  $0.463\pm0.005$  & \\
166  &  7 31 15.3  &  0.35  &  $1.63\pm0.04$  &  $0.311\pm0.005$  & \\
176  &  8 17 29.4  &  6.20  &  $2.75\pm0.05$  &  $6.588\pm0.022$  & pulsed \\
\hline
 Jan 25   &   8 43 41.6  &  0.45  & $21.68\pm0.17$  &  $3.328\pm0.020$ & \\
 \hline
\end{tabular}
\end{center}
$^{(a)}$ In the range 25 keV--2 MeV, assuming a thermal
bremsstrahlung spectrum with $kT=40$ keV.
\end{table*}

\section{Discussion}

The strong activity shown by \sgr\ on 2009 January 22, with the
emission  of hundreds of bursts in a time span of a few hours,
resembles similar episodes observed in the past from other SGRs
and AXPs (see, e.g., \citet{kas03,isr08,goe06}). The peak of the
bursting rate occurred around 6:48 UT, when more than 50 bursts
were recorded in 10 minutes. The high fluence emitted in this
short time interval is most likely responsible for the expanding
dust scattering halos discovered in \emph{Swift}/XRT  X-ray images
obtained about one day later \citep{GCN8848}. We can exclude the
presence of other periods of comparably high bursting rate in the
few preceding days, except for the time interval from 4:30 to
14:30 UT of January 20,  when INTEGRAL was at perigee.

On January 22 \sgr\ also emitted a few bursts remarkable for their
luminosity and duration, including one (n.~149) showing a bright
and short spike followed by a pulsating tail. The burst was
preceded by a short precursor, but,  considering the high rate of
bursts,  it is not clear whether this was physically related to
the bright burst.  These features of the light curve are typical
of giant flares from SGRs \citep{maz79,hur99,pal05,mer05b,hur05}.
Therefore, despite the uncertainties due to the poorly constrained
distances and possible spectral differences, it  is interesting to
compare the energetics of the \sgr\ burst n.~149 with that of the
three historical giant flares from SGRs.  Some  properties  of
such events are summarized in Table~2,  where also  a few other
peculiar bursts and flares from SGRs and AXPs are listed for
comparison.

It is evident that the energy released in the pulsating tails of
the giant flares of SGR\,0526--66, SGR\,1900+14, and SGR\,1806--20
was much higher than the value of a few 10$^{43}$ erg  we derived
for \sgr\ (unless its distance is much larger than 10 kpc).
However, the difference is mainly due to the shorter duration of
the \sgr\ pulsed tail. In fact the tails following the three
historical giant flares lasted a few minutes, but their average
luminosity ($\sim$ $3\times10^{41}$--10$^{43}$ erg s$^{-1}$) was
not too different from that observed in the pulsating tail of
\sgr\ ($\sim$ $3\times10^{42}\ d_{10\,\emph{kpc}}^{2}$ erg
s$^{-1}$).

Comparison of the initial spikes' energetics is more difficult,
since the effects of saturation and spectral uncertainties could
be larger. However, also in this case, it is clear that the
initial spike of burst n.~149 involved a significantly smaller
energy  than those of the giant flares. From the point of view of
the total energy, this event was more similar to the intermediate
flares or to some of the  brightest bursts observed in other SGRs,
such as the 1998 October 28 and the 2001 July 2 events from SGR
1900$+$14, but owing to its shorter duration it reached a higher
peak luminosity. Another notable difference with respect to the
three giant flares is that the latter had very short rise times
\citep{maz99b,sch05}, while the initial spike of burst n.~149
showed a slow rise to the peak, that could be resolved in 4 ACS
time bins (50 ms each). We note that the characteristic
time-scales over which bursts develop can give information on the
location of the energy source and on the mechanisms responsible
for their triggering \citep{tho95,lyu06a}. Instabilities in the
magnetosphere, where the Alfv\'{e}n velocity is close to the speed
of light, can develop on short timescales. Longer timescales are
expected for the release of energy stored in the neutron-star
crust.

The differences mentioned above suggest that burst n.~149 is not
of the same nature of the giant flares. Most likely this event,
and the other bright bursts observed from \sgr , are just
``normal" SGR bursts with extreme properties.  This is  also
supported by the fact that their peak fluxes and fluences are in
agreement with the distributions of these quantities derived from
the other bursts.

\begin{table*}[htbp!]
\begin{center}
\caption{Main properties of giant flares and other prominent
bursts from SGRs and AXPs.\label{flares}}
\begin{tabular}{|c|c|l|c|r|c|c|}
\hline
Source & Fluence         &  Notes$^{(a)}$ & Duration &   Energy$^{(b)}$ &   Spectrum$^{(c)}$ & Ref.$^{(d)}$ \\
Date   & (erg cm$^{-2}$) &             & (s)          & (erg)       &  (keV)   &   \\
\hline \hline
 \multicolumn{7}{|l|}{\sgr\   ($d=10$ kpc)} \\
 \hline
2009-01-22 &  $>4.6\times 10^{-5}$ ($>$25 keV) & 121-B  & 1.45 &  $>4\times 10^{42}$    &  $kT\sim40$$^{(e)}$   &    \\
 \hline
           &   $>$$1.8\times 10^{-5}$ ($>$25 keV) & 149-IHS & 0.3 &    $>$$5 \times10^{41}$ &  $kT\sim400$$^{(e)}$ &    \\
           &   $2.5\times 10^{-4}$ ($>$25 keV)& 149-PT  & $\sim$8 &  $2\times 10^{43}$   &  $kT\sim40$$^{(e)}$   &    \\
 \hline
           &  $6.6\times 10^{-5}$ ($>$25 keV)    & 176-PT  & 6.2 &  $6\times 10^{42}$     &  $kT\sim40$$^{(e)}$   &    \\
\hline \hline
 \multicolumn{7}{|l|}{SGR\,0526--66   ($d=55$ kpc)} \\
 \hline
1979-03-05 &  $4.5\times 10^{-4}$ ($>$30 keV) & IHS & 0.25 &    $3\times 10^{44}$  & $kT\sim500$  & M99   \\
           &   10$^{-3}$ ($>$30 keV)    & PT  & $\sim$200 & $3 \times10^{45}$ & $kT\sim30$    &    \\
\hline \hline
 \multicolumn{7}{|l|}{SGR\,1900+14  ($d=15$ kpc)}\\
 \hline
1998-08-27 & $>$$5.5\times 10^{-3}$ ($>$15 keV) & IHS & 0.35   & $>$$3\times 10^{44}$ &   $kT\sim300$  &   M99 \\
           & $4.2\times 10^{-3}$ ($>$15 keV)    & PT & $\sim$400 & $2\times 10^{45}$ &  $ kT\sim20$   &     \\
 \hline
1998-08-29   &   $1.9\times 10^{-5}$ ($>$25 keV) & B  & 3.5   & $2 \times10^{42}$   &  $kT\sim16$    & I01   \\
             &   $2.5\times 10^{-7}$ (3--30 keV)  & PT & $\sim$1000 & $6\times 10^{39}$ & $kT\sim100$--5$^{(f)}$  &    \\
 \hline
1998-10-28  &   $4.8\times   10^{-5}$ ($>$15 keV) & B  &  4  & $2\times 10^{41}$    & $kT\sim22$  & A01 \\
 \hline
2001-04-18    &  $1.2\times 10^{-4}$ (40--700 keV) & PT & 40 &  $3 \times10^{43}$ &   $kT\sim30$   &  G04  \\
 \hline
2001-04-28  & $8.7\times 10^{-6}$ (25--100 keV) & B &  2  & 10$^{42}$    &  $kT\sim21$   &  L03  \\
          &   $3 \times10^{-7}$ (2--20 keV)   & PT & $>$3500  & $8\times 10^{39}$   &  $kT\sim10$   &  \\
 \hline
2001-07-02     &  $1.9\times 10^{-5}$ (2--150 keV) & B  &   3.5  & $5\times 10^{41}$ &  $kT_{\mathrm{BB}}^{(g)}=5+10$    & O04   \\
\hline \hline
 \multicolumn{7}{|l|}{SGR\,1806--20 ($d=15$ kpc)} \\
 \hline
2004-12-27 &  1.4 ($>$30 keV)      &  IHS & 0.2  &   $5\times 10^{46}$   &  $kT_{\mathrm{BB}}=175$   &   H05 \\
           & $5.5\times 10^{-3}$ (3--100 keV) & PT & 380  & $1.3\times 10^{44}$     &  $kT\sim15$--30  &    \\
\hline \hline
 \multicolumn{7}{|l|}{SGR\,1627--41 ($d=11$ kpc)} \\
 \hline
 1998-06-18  &    $7 \times10^{-4}$  ($>$15 keV) & B  &  0.5 & $2\times 10^{44}$   &  $kT\sim150$   &   M99B \\
\hline \hline
 \multicolumn{7}{|l|}{XTE\,J1810--197 ($d=3$ kpc)} \\
 \hline
2004-02-16   &  $4\times 10^{-8}$  ($>$2 keV)  & B &   $\sim$0.3  &  $4\times 10^{37}$  &  $kT_{\mathrm{BB}}=7$    & W05   \\
             &  $1.4 \times10^{-7}$ ($>$2 keV) &  PT & $>$575  &  $1.4\times 10^{38}$   &  $kT_{\mathrm{BB}}=2.6$     &    \\
\hline
\end{tabular}
\end{center}

$^{(a)}$ IHS = Initial Hard Spike,  PT = Pulsed Tail, B=burst.

$^{(b)}$ In the range E$>$3 keV, assuming the indicated distances,
and isotropic emission (when needed, the spectrum has been
extrapolated to lower energy).

$^{(c)}$ Temperature of bremsstrahlung ($kT$) or blackbody
($kT_{\mathrm{BB}}$) spectra.

$^{(d)}$  References:  A01: \citet{apt01}; G04: \citet{gui04};
H05: \citet{hur05}; I01: \citet{ibr01}; L03: \citet{len03}; M99:
\citet{maz99b}; M99B: \citet{maz99c}; O04: \citet{oli04}; W05:
\citet{woo05}.

$^{(e)}$ Assumed value.

$^{(f)}$ Strong hard to soft spectral evolution.

$^{(g)}$ Fit with two blackbodies.

\end{table*}

The second pulsed burst from \sgr\ (n.~176)  did not start with a
bright  short spike. The lack of such a feature could be related
to the lower fluence of this burst (a factor $\sim$4 smaller than
that of burst n.~149) or simply to the initial spike being beamed
in a different direction from ours.  A similar situation occurred
in the flare observed  from SGR\,1900$+$14 on 2001 April 18
\citep{gui04}. It is not surprising that pulsations are seen in
all the bursts with a sufficiently long duration. \sgr\ has the
shortest period among SGRs and AXPs, and long bursts are rare. In
other objects of this class pulsating tails could only be seen
after particularly energetic events.  The pulsating tails in \sgr\
could be due to magnetically trapped fireballs, similar to the
case of giant flares, even if the mechanism responsible for the
initial energy injection is different. Alternatively, they could
result from localized regions on the neutron-star crust, possibly
heated by the same mechanisms that produced the burst. In the
latter case their modulation would not be expected to maintain
phase coherence across different bursts. We could not find strong
evidence for a phase difference between the pulses of bursts
n.~149 and n.~176, based on a backward extrapolation of the
preliminary timing solution derived with \emph{Swift}/XRT (Israel
et al., in preparation).

\section{Conclusions}

Thanks to the high sensitivity and uninterrupted coverage provided
by the SPI-ACS instrument, we could characterize the statistical
properties of the bursts emitted by \sgr\ in the few hours when
the source displayed  the most pronounced activity  during its
recent reactivation. The source is still active
\citep{GCN8913,GCN8915}, but with a much lower bursting rate
compared to that seen on 2009 January 22.

We showed that, despite the apparent similarity to giant flares,
the bright and pulsed bursts observed from \sgr\ involved a
smaller energy and are most likely related to ordinary bursts. The
total fluence measured from the 125 bursts emitted from 4:30 to
7:00 UT is $5.2\times10^{-4}$ erg cm$^{-2}$ (25 keV--2 MeV). A
fraction of the soft X-rays associated with these bursts was
forward-scattered along a somewhat longer path by interstellar
dust grains, and detected in the form of expanding rings by
several X-ray satellites about a day later. Detailed modelling of
these data will allow us to constrain the distance of \sgr\
(Tiengo et al., in preparation).

The properties of the bursts from  \sgr\ are typical of sources
classified as SGRs. Indeed, if this AXP had not been previously
known from X-ray and radio observations, it would have been named
as a new SGR following the January 22 bursts.  This underlines
once more that the distinction between these two classes of
neutrons stars is not based on physical properties of the sources,
which are most likely explained by the same model.

\acknowledgements \emph{INTEGRAL} is an ESA project with
instruments and Science Data Centre funded by ESA member states
(especially the PI countries: Denmark, France, Germany, Italy,
Switzerland, Spain), Czech Republic, and Poland, and with the
participation of Russia and the USA. The SPI anticoincidence
system is supported by the German government through DLR grant
50.0G.9503.0.  The Italian authors acknowledge the partial support
from ASI (ASI/INAF contracts I/088/06/0 and AAE~TH-058).  PE
thanks the Osio Sotto city council for support with a G.~Petrocchi
fellowship. D.G. acknowledges the CNES for financial support. N.R.
is supported by an NWO Veni Fellowship. S.Z. acknowledges support
from STFC.


\end{document}